\documentclass[twocolumn,secnumarabic,amssymb,nobibnotes, aps, prl]{revtex4-2}

\ExplSyntaxOn
\NewDocumentCommand{\longdash}{ O{2} }
 {
  --\prg_replicate:nn { #1 - 1 } { \negthinspace -- }
 }
\ExplSyntaxOff
\usepackage{graphicx}
\usepackage{natbib}
\usepackage{url}
\usepackage{textcase}
\usepackage{bm}
\usepackage{xcolor}
\usepackage{arydshln}
\usepackage{braket}
\usepackage{svg}
\usepackage{hyperref}
\hypersetup{
  colorlinks = true,
  linkcolor = blue,
  citecolor = blue
}
\usepackage{comment}
\usepackage{soul}

\def \bii{BiI$_{3}$~}
\def \Bii{BiI$_{3}$}
\usepackage{amsmath}

\begin{document}

\title{Physics of excitons in layered BiI$_3$. Effects of dimensionality and crystal anisotropy}

\author{Jorge Cervantes-Villanueva$^1$}

\email[corresponding author: ]{jorge.cervantes@uv.es}

\author{Fulvio Paleari$^2$} 

\author{Alberto Garc\'ia-Crist\'obal$^1$}

\author{Davide Sangalli$^3$}

\author{Alejandro Molina-S\'{a}nchez$^1$}
 
\address{$^1$Institute of Materials Science (ICMUV), University of Valencia,  Catedr\'{a}tico Beltr\'{a}n 2,  E-46980,  Valencia,  Spain}
\address{$^2$Centro S3, CNR-Istituto Nanoscienze, I-41125 Modena, Italy}
\address{$^3$Istituto di Struttura della Materia-CNR (ISM-CNR), Area della Ricerca di Roma 1, Monterotondo Scalo, Italy}

\date{\today}

\begin{abstract}

We carry out a detailed theoretical study of the electronic and optical properties of bulk and monolayer bismuth triiodide (\Bii), a layered metal halide, using the \textit{ab initio} GW+BSE scheme with a full spinorial formulation. We discuss in detail the effects due to the change of dimensionality and the role of spin-orbit coupling. Moreover, we compute the exciton dispersion by solving the BSE at finite momentum, also analysing transverse (TE) and longitudinal (LE) excitons, and the L-T splitting at $\bf{q}\approx\Gamma$. The results provide a reference for future experimental measurements. In addition, the interplay between spin-orbit coupling and large binding energy, together with the role of quantum confinement, confirm that \bii is an interesting material for opto-electronic applications and show that it is a good candidate for the study of exciton dynamics.

\end{abstract}

\maketitle

\section{Introduction}

Exciton physics is a fascinating topic of research in condensed matter, displaying phenomena such as Bose Einstein condensation (BEC), \cite{moskalenko_snoke_2000, morita2022observation} topological non-equilibrium phases\cite{Perfetto2020}, BEC-Bardeen Cooper Schrieffer (BCS) crossover \cite{Perfetto2021BECBCS}, the exciton-Mott transition\cite{chernikov2015population}, with especially remarkable properties in the realm of two-dimensional materials\cite{Wang2018}. The experimental investigation found an ideal playground in two-dimensional (2D) materials~\cite{mueller_exciton_2018}, such as transition metal dichalcogenides (TMDs), which present large excitonic binding energies of around 500 meV~\cite{ugeda_giant_2014}. The large binding energy and the optical bandgap in the visible range make TMDs promising materials for optoelectronic device applications, such as electrically driven light emitters, photovoltaic solar cells, photodetectors, and valleytronic devices \cite{wang_electronics_2012,schaibley_valleytronics_2016}. Recent experiments demonstrate the electrical control of excitons \cite{tagarelli_electrical_2023} and the properties of exciton complexes \cite{pasquale_flat-band-induced_2022}, along with promising application of 2D materials in solid-state devices to implement spin-based computation schemes or to investigate bosonic interactions \cite{ciarrocchi_excitonic_2022}.

In particular, time-resolved spectroscopy is providing unprecedented understanding of exciton properties and dynamics. Recent experiments using time-resolved angle-resolved photoemission spectroscopy (tr-ARPES) \cite{madeo_directly_2020} have succeeded in probing exciton dynamics in the femto-second regime, boosting the interest in exciton physics. Tr-ARPES has been proven useful to monitor the formation of moir\'e interlayer excitons in TMDs heterobilayers \cite{schmitt_formation_2022,Torun2018}, the hot electron dynamics in monolayer MoS$_2$ \cite{lee_time-resolved_2021}, and direct imaging of excitonic wavefunctions \cite{man_experimental_2021}. 

The materials suitable for exploring the physics and dynamics of excitons exhibit high excitonic binding energies and large area samples. Mechanical exfoliation of layered materials achieves a low number of layers, enhancing the excitonic binding energy via quantum confinement and reduced electronic screening of the excitations \cite{raja_coulomb_2017}. In this context, the layered metal halide bismuth triiodide (\Bii) is a promising candidate for exploring physics of excitons using experimental techniques like tr-ARPES. Bulk \bii is a semiconductor 
with strong absorption in the visible range, becoming an important candidate for photovoltaic applications \cite{brandt2015investigation, lehner2015electronic} and for room-temperature gamma-ray detectors
\cite{matsumoto2002bismuth, rosztoczy1965bismuth}. Initial experimental studies have shown that the bulk has an exciton binding energy of approximately 180 meV, which is unusually large \cite{kaifu1988excitons,Mor2021}. 

The \bii bandgap value has also been reported as a function of the number of layers, indicating a large difference between the optical and electronic bandgaps \cite{mu_resolving_2021}. Plates and thin films of \bii have been obtained by different methods such as hot wall techniques and thermal evaporation methods \cite{takeyama1990thin, garg2014synthesis}. Monolayer \bii has not yet been obtained; however, according to a recent theoretical work \cite{ma2015single}, mechanical exfoliation of bulk \bii to fabricate monolayers should be possible as in the paradigmatic 2D semiconductor monolayer MoS$_{2}$. In addition, bismuth and iodine have large atomic numbers and thus \bii is subject to a strong spin-orbit interaction, with potential interest for spin physics in 2D materials \cite{ahn_2d_2020}.

Despite the promising optical properties of bulk \bii and the possibilities of synthesizing monolayers, to our knowledge there are no theoretical or computational studies analysing the physics of excitons in \Bii. So far, research has focused on the electronic properties \cite{lehner2015electronic}, optical properties without including spin-orbit coupling \cite{xiao2021large}, and, more recently, a study on coherent phonons \cite{Mor2021}. In this context, a complete study of the excitonic states of bulk and monolayer \Bii, together with the exciton dispersion, is a highly needed reference point. In this work we characterise the excitonic states of bulk and monolayer \Bii, the two opposite cases with respect to the magnitude of quantum confinement effects and dielectric screening. We start with a comparative analysis of the electronic structure within the GW approach, which is achieved by projecting the bulk band structure onto the 2D Brillouin zone (BZ) of the monolayer. Then we diagonalize the excitonic hamiltonian, analyse bright and dark excitons, and compute the optical properties of the material. We also compute the exciton dispersion, carefully analyse the longitudinal-transverse (L-T) splitting of bright excitons and compare our results with the L-T splitting measured by reflection spectroscopy in the bulk \cite{kaifu1988excitons}.

The paper is organized as follows. In section~\ref{theory} we report the computational details of the \textit{ab initio} GW+BSE scheme. In section~\ref{results} we present our results for the electronic and optical properties of bulk and monolayer \Bii. The physics of excitons in the two systems is investigated in detail, both at zero momentum $\bf{q}=\Gamma$, and by analysing the excitonic dispersion together with the L-T splitting. In section~\ref{discuss} we present the conclusions.

\begin{figure*}
\includegraphics[width=1.00\linewidth]{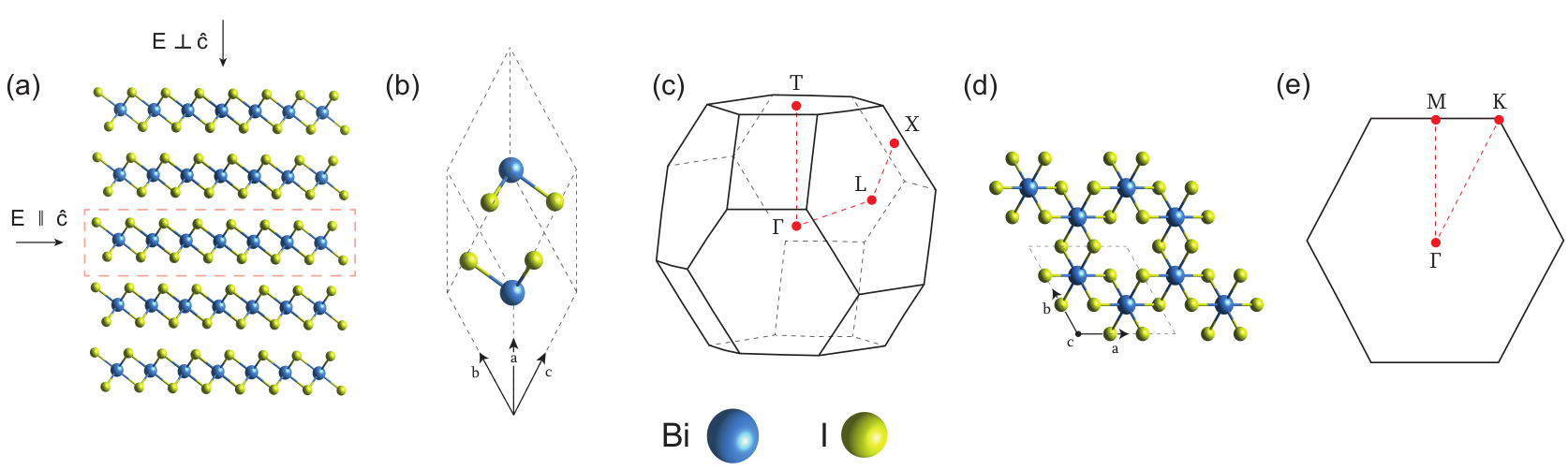}
\caption{\label{Estructure} Crystal structure and Brillouin zone of bulk and monolayer \Bii. (a) Side view of bulk \Bii, where the dashed box represent the monolayer \Bii, (b) unit cell of bulk \Bii, (c) Brillouin zone of bulk \Bii, (d) unit cell of monolayer \Bii, (e) Brillouin zone of monolayer \Bii.}
\end{figure*}

\section{Theory and computational details\label{theory}}

Our theoretical study of the exciton physics in \bii is based on first-principles calculations. The computation of the electronic properties is carried out within the framework of density functional theory (DFT), as implemented in the Quantum ESPRESSO package \cite{giannozzi2009quantum,giannozzi2017advanced,giannozzi2020quantum}. We have employed
the Perdew–Burke–Ernzerhof (PBE) exchange-correlation functional \cite{perdew1996generalized}. \bii is a layered metal halide with a highly ionic bond between bismuth and iodine atoms within the same layer, and weak van der Waals forces between layers. Accordingly, we account for van der Waals interaction with the Tkatchencko-Scheffler model (TS-vdW)~\cite{PhysRevLett.102.073005}. Due to the large atomic number of bismuth and iodine atoms, the spin-orbit coupling dominates the electronic properties and therefore fully relativistic norm-conserving pseudopotentials are used from PseudoDojo \cite{van2018pseudodojo}. 
The effect of the semi-core electrons is considered including explicitly in valence the $5d$ semi-core states for iodine and $5d$ semi-core states for bismuth, due to their reported importance in the electronic structure calculations \cite{scherpelz2016implementation, giustino2014materials}. In previous computational works optimized lattice parameters were used, which however differ by more than 4\% with respect to those extracted from experimental data \cite{Mor2021}. In the present work we use the experimental lattice parameters of the bulk \cite{nason1995growth} for both bulk and monolayer (see Table \ref{ComputationalValues}). In the case of the monolayer, a 15 \r{A} vacuum thickness is imposed in all calculations to avoid fictitious interactions due to the supercell periodicity. The calculations at the DFT level are carried out with a plane-wave cutoff of 80 Ry and the \textbf{k}-grid reported in Table \ref{ComputationalValues} following the Monkhorst-Pack method \cite{monkhorst1976special}.

\begin{table}[!t]
\caption{\label{ComputationalValues} Computational parameters for the calculation of the electronic and optical properties of bulk and monolayer \Bii.}
\begin{ruledtabular}
\begin{tabular}{ccccc}
 &\multicolumn{2}{c}{\textbf{Bulk}}&\multicolumn{2}{c}{\textbf{Monolayer}}\\ \hline
a &\multicolumn{2}{c}{7.52 \r{A}}&\multicolumn{2}{c}{7.52 \r{A}}\\
c &\multicolumn{2}{c}{20.7 \r{A}}&\multicolumn{2}{c}{\longdash[3]}\\
GW bands &\multicolumn{2}{c}{300}&\multicolumn{2}{c}{300}\\
Static $\epsilon (\omega=0)$ bands &\multicolumn{2}{c}{200}&\multicolumn{2}{c}{250}\\
BSE bands &\multicolumn{2}{c}{139 - 152}&\multicolumn{2}{c}{134 - 152}\\
\textbf{k}/\textbf{q}-grid &\multicolumn{2}{c}{8$\times$8$\times$8}&\multicolumn{2}{c}{18$\times$18$\times$1}\\
\end{tabular}
\end{ruledtabular}
\end{table}

The inherent underestimation of the DFT bandgap is corrected using the GW method \cite{reining2018gw, onida2002electronic}, while optical and excitonic properties are obtained solving the Bethe-Salpeter equation \cite{strinati1988application}, both in the framework of many-body perturbation theory (MBPT) as implemented in the Yambo code \cite{marini_yambo_2009,sangalli2019many, marsili2021spinorial}. Quasiparticle corrections of the DFT eigenvalues are introduced in a single-shot GW approach (G$_{0}$W$_{0}$) \cite{stan2009levels}, using the plasmon-pole approximation for the dynamical dependence of the dielectric function \cite{aryasetiawan1998gw}. The dielectric function is calculated in the random phase approximation \cite{RPA}.  A truncated Coulomb potential in a slab geometry is used in the monolayer case. This is necessary to avoid interactions between periodic images and to accelerate the convergence of the GW calculations with respect to the k-points \cite{guandalini2022efficient}. The number of bands included in the calculation of the dielectric matrix in the GW calculation is reported in Table \ref{ComputationalValues}. In addition, a terminator
is used to speed up the convergence of the electronic Green's function $G$ with respect to the empty states \cite{bruneval2008accurate}. For the study of the optical properties taking into account excitonic effects, the BSE is solved on top of the G$_{0}$W$_{0}$ results within the Tamm-Dancoff approximation \cite{strinati1988application}. The exciton dispersion is calculated by solving the finite momentum BSE in the transferred momenta $\mathbf{q}$-grid shown in Table \ref{ComputationalValues}.
The number of bands used to build the statically screened kernel of the BSE and the valence and conduction bands involved in the calculation of the excitonic transitions are reported in Table \ref{ComputationalValues} as well. Due to the anisotropy of the system, both for GW and BSE calculations the average of ${\bf E} \perp \hat{c}$ and ${\bf E} \parallel \hat{c}$ polarizations is adopted to deal with screening at $\bf{q}=\Gamma$, $\hat{c}$ being the stacking direction.

\section{Results}\label{results}

\subsection*{Electronic properties}

The crystalline structure of bulk and monolayer \bii is shown in Fig.~\hyperref[Estructure]{1(a)}. The bulk crystal structure belongs to the space group R-3 (No. 148) \cite{ruck1995darstellung} and double point group symmetry S$_{6}$. The system is defined by a rhombohedral unit cell made of six iodine atoms and two bismuth atoms, as depicted in Fig.~\hyperref[Estructure]{1(b)}. This results in a three-layer packing structure characterised by an ABC stacking \cite{yorikawa2008theoretical}. On the other side, the crystal structure of the monolayer belongs to the space group P-31m (No. 162) \cite{zhou2023two} and double point symmetry D$_{3d}$. The monolayer is determined by an hexagonal unit cell of six bismuth and two iodine atoms, similar to the bulk, Fig.~\hyperref[Estructure]{1(d)}, with the characteristic hollow of trihalides like CrI$_{3}$ \cite{huang2017layer}. As we will see in the following, this particular stoichiometry brings the structure of this solid closer to that of a molecular crystal with respect to other layered systems, with important consequences on the optical and excitonic properties.

\begin{figure*}[t!]
\includegraphics[width=1.0\linewidth]{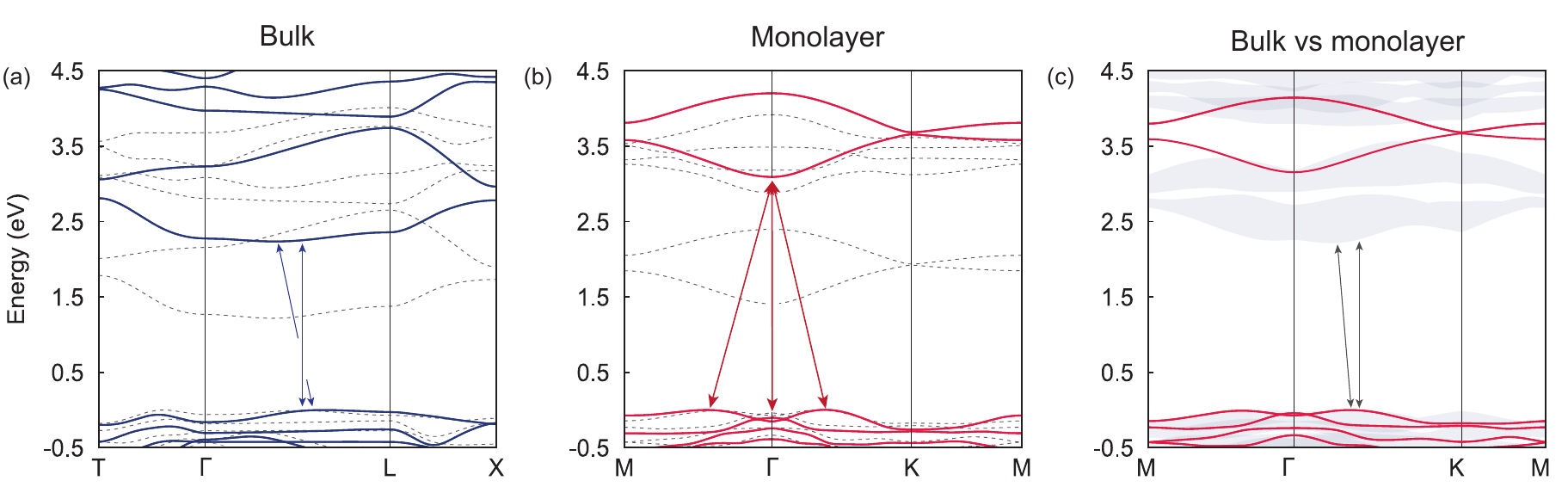}
\caption{\label{Bstructure} Electronic band structures of (a) bulk and (b) monolayer \Bii. The solid lines represent the GW bands while the dashed ones represent the DFT ones. (c) DFT band structure of bulk (shaded areas) and monolayer (red lines) \Bii, both represented  in the hexagonal Brillouin zone. The shaded areas represent the integrated dispersion over $k_{z}$ of the bulk.  A scissor operator extracted from the GW calculations is applied to open the gaps.}
\end{figure*}

We next analyse the electronic band structure of bulk and monolayer \bii as obtained from calculations at DFT and GW levels. The bulk and monolayer band structures are shown in Fig.~\hyperref[Bstructure]{2(a,b)}, 
respectively.
The paths in the corresponding  BZs are indicated in Fig.~\hyperref[Estructure]{1(c,e)}.
For clarity, we have selected a simplified path for the bulk case (a more detailed band structure can be found in Appendix \ref{FullBSBiI3}). 
The lowest direct and indirect gaps are indicated by arrows
in the corresponding band structures. 
In both cases the fundamental gap is found to be indirect, which is consistent with the conclusions reported previously in the literature \cite{podraza2013band, xiao2021large}, although the difference with the direct bandgap is rather small.
To better understand how the change in dimensionality affects the electronic properties, the band structure of monolayer (red lines) and bulk (shaded gray area) \bii are shown together in Fig.~\hyperref[Bstructure]{2(c)} in the planar hexagonal BZ. Here, the bands of the bulk are integrated over $k_{z}$. The region around $\Gamma$ is the one defining the bandgap in both cases. The characteristic off-$\Gamma$ valence band maxima (VBM) are present both in monolayer and bulk. However, the conduction band minimum (CBM) of the monolayer is located at $\Gamma$, whereas in the bulk it is located along the line from $\Gamma$ to $K$. This means that, when going from bulk to monolayer, the minimum of the CBM is shifted, slightly changing the wave vector associated to the indirect gap as well as the position of the direct one, analogously to what is observed in TMDs \cite{mak2010atomically,molina-sanchez_vibrational_2015}. The direct and indirect gaps of the bulk in the hexagonal BZ are marked by arrows.

\begin{table}[!b]
\caption{\label{GapValues} Bandgap energy values for bulk and monolayer \bii calculated at DFT and GW levels.}
\begin{ruledtabular}
\begin{tabular}{ccccc}
 &\multicolumn{2}{c}{DFT}&\multicolumn{2}{c}{{GW}}\\
  & {Direct} & {Indirect} & {Direct}
& {Indirect} \\ \hline
 {Bulk} & 1.24 & 1.19   & 2.26 & 2.24 \\
 {Monolayer} & 1.45 & 1.41   & 3.19 & 3.09 \\
\end{tabular}
\end{ruledtabular}
\end{table}

The calculated bandgap values are reported in Table \ref{GapValues}. They can be compared with the experimental bandgap values of 2.65 eV (bulk) and 2.8 (monolayer) measured recently by scanning tunneling spectroscopy \cite{mu_resolving_2021}. Nevertheless, these experimental values can strongly depend on the sample quality.
The difference between the direct and indirect GW bandgaps for the bulk is 20 meV, the bulk exhibiting a much flatter conduction band than in the case of the monolayer. This subtle difference between the indirect and direct bandgap might be the reason of the discrepancies in the reported values of the fundamental gap and its character (direct or indirect) as extracted from absorption measurements~\cite{podraza2013band}.
Our results confirm that the spin-orbit coupling  plays a crucial role in describing the electronic properties of this material since it strongly shifts the CBM down, decreasing the gap value by approximately 1.0 eV (see Appendix \ref{BSnosocBiI3} 
where a comparison is made between the band structures obtained with and without spin-orbit interaction).

In addition, if the $5d$ semicore orbitals for iodine are not included in the DFT+GW calculation, the quasiparticle corrections are wrong by up to 0.5 eV. For both bulk and monolayer crystals the quasiparticle corrections significantly modify the valence bands, yet essentially by producing a rigid shift of the conduction bands. In any case, they do not change the nature of the bandgaps. The bandgap renormalization due to the quasiparticle corrections is approximately 1.0 eV and 1.8 eV for bulk and monolayer, respectively. The large difference between bulk and monolayer is linked to quantum confinement effects and to the much weaker dielectric screening in the planar monolayer, which enhances the Coulomb interaction \cite{chernikov2014exciton}. Moreover, the peculiar crystalline structure of \Bii, with a hollow site in the lattice formed by bismuth atoms, further reduces the screening, and is also responsible of the large bandgap renormalization exhibited by bulk \Bii.

Concerning other \textit{ab initio} calculations, a direct bandgap value of 2.10 eV was found 
for the bulk system in a recent work \cite{Mor2021} using the GW approach, but without including the $5d$ semi-core states for iodine atoms and with lattice parameters considerably different from the experimental ones. In the case of the monolayer crystal, a GW direct bandgap of 4.20 eV was recently reported in the literature~\cite{xiao2021large}, however without considering spin-orbit coupling.  

\subsection*{Exciton Physics}

\begin{figure*}
	\includegraphics[width=1.0\linewidth]{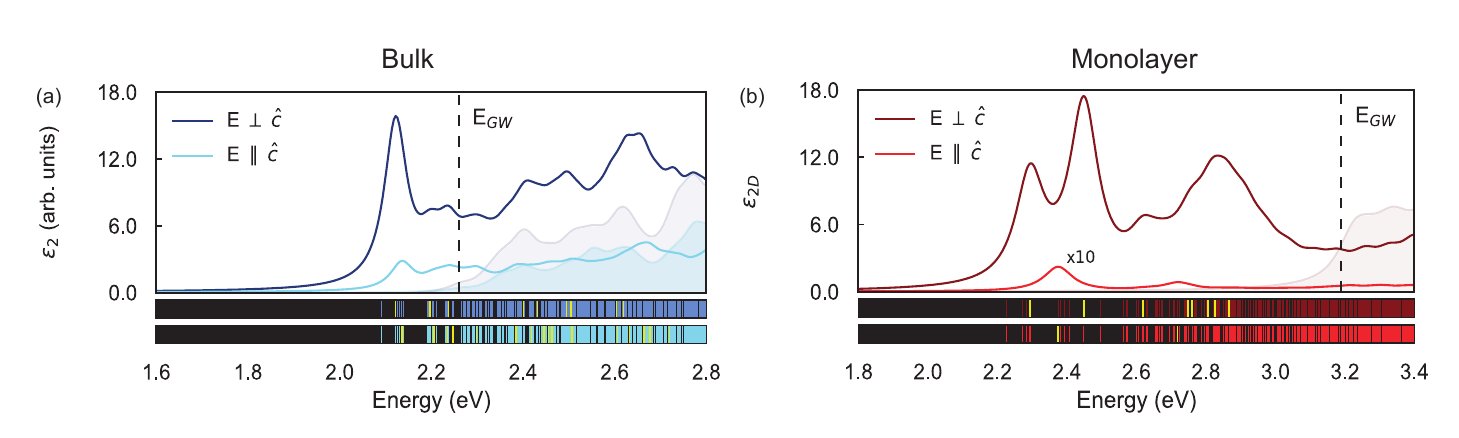}
	\caption{\label{Oabsorption} Absorption spectra for (a) bulk and (b) monolayer \bii obtained with the IP-RPA  (shaded areas) and BSE (solid lines) for ${\bf E} \perp \hat{c}$ and ${\bf E} \parallel \hat{c}$ polarizations. In the case of the monolayer, the BSE spectrum for the ${\bf E} \parallel \hat{c}$ polarization has been multiplied by a factor of 10 to be easily observed. The GW direct bandgap is marked with a vertical black dashed line. In the insets below, the yellow vertical lines indicate the bright excitons and the rest the dark excitons.}
\end{figure*}

The excitonic effects and the optical properties are then computed on top of the quasiparticle band structure. The absorption spectra of the bulk and monolayer \bii for ${\bf E} \perp \hat{c}$ and ${\bf E} \parallel \hat{c}$ polarizations are shown in Fig.~\hyperref[Oabsorption]{3(a,b)}, where GW+BSE results are compared to the optical absorption without excitonic effects (shaded area), calculated in the independent particle random-phase approximation (IP-RPA). The GW direct bandgap energy is represented by a vertical dashed line, while the excitonic transitions are represented in the insets below with vertical lines. The yellow lines represent the bright excitons and the blue/red lines the dark excitons. In the bulk case shown in Fig.~\hyperref[Oabsorption]{3(a)}, the optical absorption refers to the imaginary part of the macroscopic dielectric function obtained in the calculations, $\epsilon_2$. In  the monolayer case depicted in Fig.~\hyperref[Oabsorption]{3(b)}, the imaginary part of the macroscopic dielectric function is not well defined and is more appropiate to work with the polarizability per unit area, $\alpha_{2D}(\omega)$. In order to compare with the bulk case, we define a 2D effective dielectric tensor \cite{molina2020magneto}: 

\begin{equation}
	\epsilon_{2D}(\omega) = 1 + \frac{4 \pi \alpha_{2D}(\omega)}{\Delta z}    
\end{equation}

with $\Delta z$ the material thickness in the hexagonal unit cell, being in this case $\Delta z = 7.82$ \r{A}. In both cases the optical response is larger for ${\bf E} \perp \hat{c}$ polarization (usually referred to as ordinary geometry) than for ${\bf E} \parallel \hat{c}$ (extraordinary geometry). This is especially so in the monolayer, due to the smaller cross section area. For that reason from now on we will consider the ${\bf E} \perp \hat{c}$ case as a reference when necessary. 

In the bulk material the first bright exciton transition for ${\bf E} \perp \hat{c}$ (${\bf E} \parallel \hat{c}$) is located at 2.12 eV (2.14 eV) therefore indicating an exciton binding energy of 140 meV (120 meV). Experimental measurements on bulk samples reported in the literature \cite{kaifu1988excitons} indicated an excitonic peak at 2.07 eV (with an exciton binding energy of 180 meV) for ${\bf E} \perp \hat{c}$, and at 2.08 eV for ${\bf E} \parallel \hat{c}$, both at a temperature of 2K. Other measurements show that the value of the excitonic peak decreases, when increasing temperature, down to 0.1 eV \cite{komatsu1976optical}. Given that neither temperature effects nor the shifts induced by the zero point motion of ions are considered in our calculations, we consider the theoretical values, although slightly larger, in good agreement with the low-temperature experimental ones.

In the monolayer system the excitonic peak of the first bright exciton for ${\bf E} \perp \hat{c}$ (${\bf E} \parallel \hat{c}$) is at 2.30 eV (2.38 eV) with an exciton binding energy of 893 meV (810 meV). The increase of the binding energy compared to the bulk is again due to the enhancement of both the Coulomb interaction (by the change of dielectric environment) and the quantum confinement of electrons, as previously mentioned in the context of the GW calculations. In the previously reported \textit{ab initio} calculations for the monolayer \cite{xiao2021large}, a larger value of 3.18 eV was obtained for the first bright exciton transition for ${\bf E} \perp \hat{c}$. The discrepancy with our results is likely due to the neglect of the spin-orbit coupling in Ref. \cite{xiao2021large}. 

Remarkably, the values of the exciton binding energies are larger than those of most TMDs. Indeed, for the bulk case~\cite{jung2022unusually} only MoTe$_{2}$ has a similar binding energy of 150-160 meV~\cite{robert2016excitonic, arora2017interlayer}.
The exciton binding energies of the \bii monolayer are much larger than those of the TMDs monolayers, which range between 420 meV for WTe$_{2}$ up to 550 meV for the paradigmatic MoS$_{2}$ \cite{ugeda_giant_2014,haastrup2018computational}.
This makes \bii both a good candidate for exciton-based devices, and as a playground to study exciton dynamics at room temperature. For instance, measurements with tr-ARPES would allow for the studying of exciton dynamics even in bulk materials~\cite{sangalli2021excitons}.

\begin{figure}[!b]
	\includegraphics[width=0.75\linewidth]{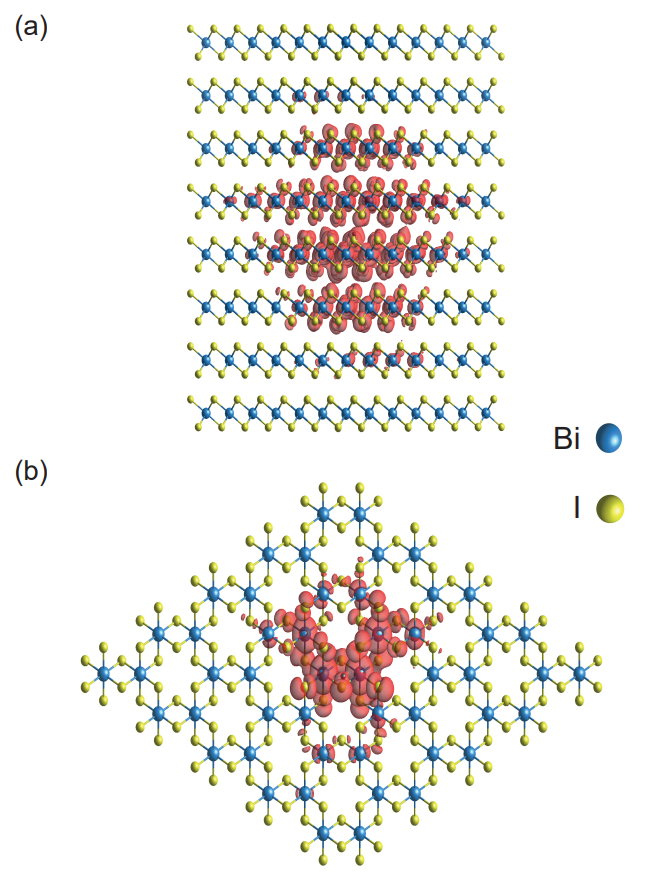}
	\caption{\label{ExcWF} Exciton wavefunction of the first bright exciton for ${\bf E} \perp \hat{c}$ for (a) bulk and (b) monolayer \Bii. The isosurface value used to depict the exciton wavefunctions is 15\% of the maximum intensity in both cases.}
\end{figure}

\begin{figure*}[!t]
	\includegraphics[width=1.0\linewidth]{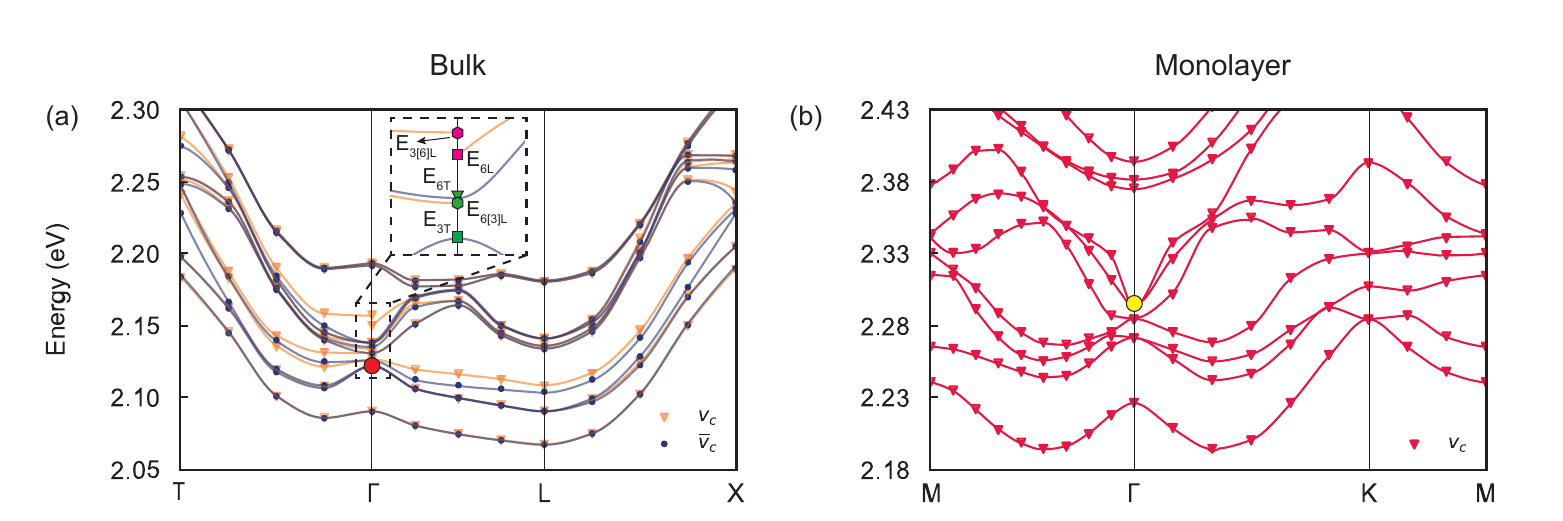}
	\caption{\label{ExcDisp} Exciton dispersion for (a) bulk and (b) monolayer \Bii. Dots are exact calculations and lines represent interpolated values. The big red and yellow dots indicate the position of the first bright exciton for ${\bf E} \perp \hat{c}$ polarization for bulk and monolayer, respectively. The curves labeled with $\bar{v}_{c}$ ($v_{c}$) have been obtained without (with) the inclusion of the long-range exchange electron-hole interaction. In (a), the marks in the inset represent the energies of the optically active excitons for the directions indicated in Fig.~\hyperref[LTsplitting]{6}.}
\end{figure*}

To gain more insight about the excitons and the origin of the large binding energies, exciton wavefunctions are shown in real space in Fig.~\ref{ExcWF}. The hole is placed on an iodine atom in both cases, since the electrons are located around it in the VBM. The exciton in the bulk is relatively localized considering that it is a bulk system. The reason for this can be partially traced to the crystal structure of the system. In addition to being a layered material, in \bii the crystalline structure is not completely homogeneous within the layer plane, but presents hollow regions that result in a stronger localization of the exciton wavefunction. This effect is more pronounced in the case of the monolayer, where there is no periodicity along the z-axis so the confinement effect is quite large. This translates into a much higher exciton binding energy \cite{Excitonlocalized}.

\subsection*{Exciton dispersion}

The indirect nature of the bandgap of bulk and monolayer \bii makes it necessary to understand the details of the exciton dispersion as a function of the center-of-mass momentum \textbf{q}  to describe its dynamics and the possibility of phonon-assisted optical transitions. For this purpose, we solve the BSE for finite momentum transfer \textbf{q} within the Brillouin zone. The BSE describes the coupling of an electronic valence-to-conduction transition ${i=\{v\mathbf{k-q}\rightarrow c\mathbf{k}\}}$ with a second transition ${j=\{v'\mathbf{p-q}\rightarrow c'\mathbf{p}\}}$, and the BSE kernel $K$ has the following structure in momentum space~\cite{marsili2021spinorial}:
\begin{equation}
-iK_{ij}(\textbf{q}) \equiv [W_{ij}^{\textrm{st}}(\textbf{k} - \textbf{p}) - V_{ij}(\textbf{q})].
\label{Eq2}
\end{equation}
In the above abbreviated expression, the first term is the statically screened electron-hole attraction $W^{\textrm{st}}$ and the second one is the bare electron-hole exchange $V$. Notice the different momentum dependence of the two terms. Eq.~\eqref{Eq2} can be solved with two forms of the exchange term $V$, either including or not the long-range contribution $v_{\mathbf{G}}^{c,LR}(\mathbf{q})$ of the bare Coulomb interaction  $v_{\mathbf{G}}^c(\mathbf{q})=4\pi/|\mathbf{q}+\mathbf{G}|^2$, here $\mathbf{G}$ being a reciprocal lattice vector. We can thus define the ``full range'' $V^{FR}_{ij}(\textbf{q}) = \langle i | v^{c}(\mathbf{q})|j \rangle$, or the ``short-range'' ${V}^{SR}_{ij}(\mathbf{q}) = \langle i | \bar{v}^{c}(\mathbf{q})|j \rangle$, where $\bar{v}^{c}(\mathbf{q})=v^{c}(\mathbf{q})-v_{\mathbf{G}}^{c,LR}(\mathbf{q})$. The inclusion of the long-range contribution leads to an important qualitative difference in the results, since it determines the splitting between transverse excitons (TE, obtained when $V^{SR}(\textbf{q})$ is used in Eq. \eqref{Eq2}) and longitudinal excitons (LE, obtained when the full $V^{FR}(\textbf{q})$ is applied)~\cite{andreani_1988}, in analogy to the LO-TO splitting of the optical phonon modes in polar semiconductors \cite{yu_fundamentals_2010}; neglecting the long-range contributions nets only the TE energies. The type of L-T splitting can be understood as a manifestation of the anisotropy of the crystal lattice in the excitonic properties. In this case, $v_{\mathbf{G}}^{c,LR}(\mathbf{q})$ plays a similar role as
the long-range dipolar electric field in the phonon case.
Indeed, for $\mathbf{q}\rightarrow 0$ ~\cite{DelSole1984,Qiu2021},
\begin{equation}\label{eq:exchange}
 \langle i | v_{\mathbf{G}}^{c,LR}(\mathbf{q}) | i \rangle  \propto \frac{(\mathbf{q}\cdot\mathbf{d}_i)^*(\mathbf{q}\cdot\mathbf{d}_i)}{q^{p-1}} =|d_i|^2 q^{3-p} cos^2(\theta)  
\end{equation}
where $\mathbf{d}_i$ is the electric dipole matrix element for the transition $i$, and $\theta$ the relative angle between $\mathbf{d}_i$ (directed along the external electric field) and $\mathbf{q}$. The presence of the dipoles implies that only optically bright excitons, for which $|\mathbf{d}_i|\neq 0$, can be subject to L-T splitting~\footnote{to prove this, it is enough to rotate $\langle i | v^{c,LR}(\mathbf{q}) | j \rangle$ in the excitonic space}. Finally, $p=2,3$ in 2D and 3D respectively. This leads to different behaviours in different dimensions~\cite{agranovich2013crystal}. In 3D we have $v_{\mathbf{G}}^{c,LR}(\mathbf{q})=v_{\mathbf{G}=0}^{c}(\mathbf{q})=4\pi/|\mathbf{q}|^2$, and there is a finite L-T splitting at $\bf{q}=\Gamma$ which depends on the direction along which the limit $\mathbf{q}\rightarrow 0$ is approached. In terms of physical observables, optical absorption -- the response of the system to a transverse electromagnetic field -- is proportional to the imaginary part of the  $\textbf{q}=0$ macroscopic transverse dielectric function $\mathrm{Im}\epsilon_T(\omega)$, and $\epsilon_T(\omega)=\epsilon_L(\omega)$ in this limit. Instead, the LE can be observed -- also at finite $\mathbf{q}$ -- alongside plasmons in electron energy loss spectroscopy (EELS), which is proportional to $\mathrm{Im}\epsilon_L^{-1}(\textbf{q},\omega)$.
In Fig. \hyperref[ExcDisp]{5(a)} we show the results of calculations performed using either $V(\textbf{q})=V^{FR}(\textbf{q})$ or $V(\textbf{q})=V^{SR}(\textbf{q})$ in Eq.~\eqref{Eq2}.
In contrast with the 3D case, in the isolated 2D system the long-range behavior of the exchange interaction emerges from the integration along the z direction of $\langle i |v_{G_x=G_y=0}^{c}(\mathbf{q})| j \rangle$ \cite{Cudazzo2016}.
This results in a term proportional to $|q|$ in the BSE Kernel (see Eq.~\eqref{eq:exchange}). Accordingly, at zero momentum there is no effect due to the long range exchange, e.g. there is no L-T splitting at $q=0$ in 2D.
At finite momentum, because of depolarization effects linked to the effectively infinite $\hat{c}$-axis of the 2D system $\epsilon_L(\textbf{q},\omega)$ and $\epsilon^{-1}_L(\textbf{q},\omega)$ capture the same physics at any $\mathbf{q}$ \cite{Hambach2010}, with poles matching the solution of Eq.~\eqref{Eq2} with the long range exchange included. For this reason in Fig.~\hyperref[ExcDisp]{5(b)} we only show the case in which $V(\textbf{q})=V^{FR}(\textbf{q})$ \footnote{Our BSE calculations are restricted to $\epsilon_L$ in this work. At finite momentum in 2D the long-range exchange still gives two branches with two different dispersions, as already discussed in the literature: one parabolic with dipole perpendicular to the $\mathbf{q}$ direction (TE branch), and one linear, with dipole parallel to the $\mathbf{q}$ direction (LE branch), e.g. there is L-T splitting at finite momentum in 2D. The TE branch could be detected by measuring $\epsilon_T(\textbf{q},\omega)\neq \epsilon_L(\textbf{q},\omega)$ at $\mathbf{q}\neq 0$.}.

The exciton dispersions of bulk and monolayer \bii are shown in Fig. \hyperref[ExcDisp]{5(a,b)} for the first 10 excitons. For both systems the minima of the lowest energy excitons are not located at $\Gamma$, which is consistent with having a fundamental indirect gap. We first focus on the bulk, Fig.~\hyperref[ExcDisp]{5(a)}, where the excitons are labeled with increasing integers starting from the lowest energy exciton. Thus, the lowest energy exciton $E_1$ exhibits its minimum at $\bf{q}=L$. $E_1$ is dark and, accordingly, not subject to L-T splitting. $E_2$ and $E_3$ are degenerate (red circle) and bright for the ${\bf E} \perp \hat{c}$ polarization. Thus, they are subject to L-T splitting, being maximum when the direction of the electric field is in-plane. Finally, $E_4$ and $E_5$ are again dark, while $E_6$ is bright for the ${\bf E} \parallel \hat{c}$ polarization. In this case $E_6$ is nondegenerate, but it is subject to an energy shift, displaying L-T behavior and the shift being maximum when the direction of the electric field is out-of-plane. The rest of the excitons are dark and therefore not subject to L-T splitting.

To analyse the nonanalytic behaviour at $\bf{q}=\Gamma$ of the previous excitons, we calculate LE considering two electric field directions in the limit $\bf{q}\rightarrow0$, e.g.  along the directions T$ \rightarrow \Gamma$ and L$ \rightarrow \Gamma$ at $\bf{q}=\Gamma$, in line with the exciton dispersion. This is depicted in Fig.~\hyperref[ExcDisp]{5(a)} by the orange lines, where the inset shows the excitons of interest. As expected, an energy shift leading to L-T splitting is only observed in the optically active excitons, which are represented by the green and magenta marks.

The exciton dispersion of the monolayer is depicted in Fig.~\hyperref[ExcDisp]{5(b)} and shows that the lowest-energy exciton has two minima matching the energy indirect transitions of the band structure calculated in Fig.~\hyperref[Bstructure]{2(b)}. In contrast to the bulk, several dark excitons lie below the first bright exciton (yellow circle) for the ${\bf E} \perp \hat{c}$ polarization. The linear dispersion can be observed when moving away from $\bf{q}=\Gamma$ for the first bright exciton, in agreement with Eq. \eqref{eq:exchange} and what has been reported in previous studies for 2D materials \cite{Cudazzo2016,Qiu2021}.

\begin{table}[!b]
\caption{\label{ExcTransDir} Energies of the excitons E$_{3}$ and E$_{6}$ when the electric field direction changes from ${\bf E} \perp \hat{c}$ to ${\bf E} \parallel \hat{c}$.}
\begin{ruledtabular}
\begin{tabular}{cccc}
 E$_{3T}$ & E$_{6T}$ & E$_{6L}$ & E$_{3L}$ \\
 2.122 & 2.136 & 2.150 & 2.160 \\
\end{tabular}
\end{ruledtabular}
\end{table}

\begin{figure}[t]	\includegraphics[width=1.0\linewidth]{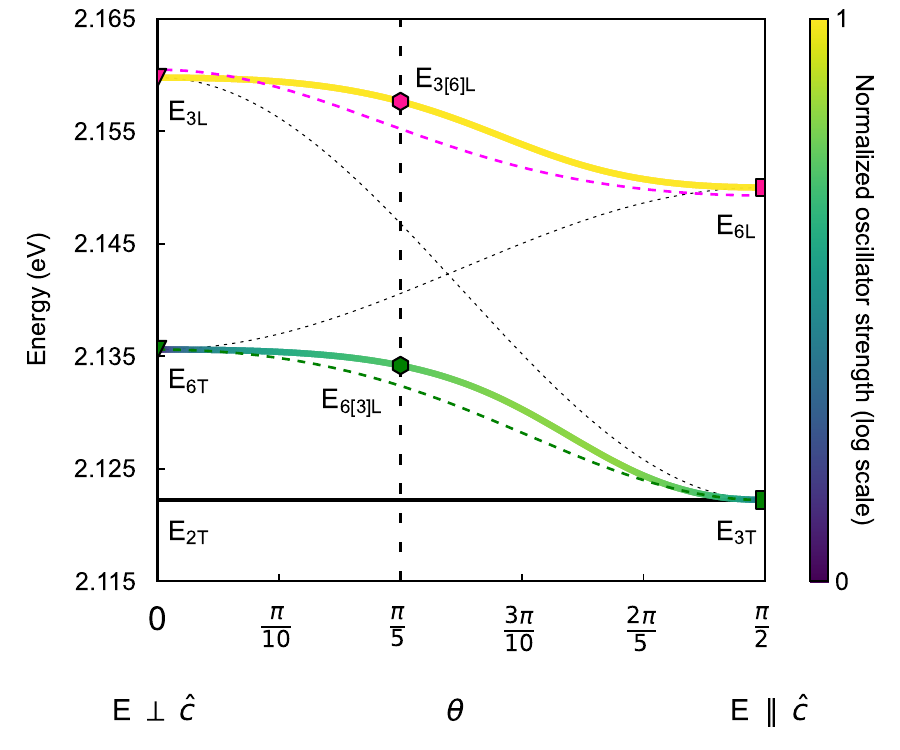}
        \label{LTsplitting}
	\caption{Behaviour of the excitons $E_{2-3}$ and E$_{6}$ at $\bf{q}=\Gamma$ when the electric field direction changes from ${\bf E} \perp \hat{c}$ to ${\bf E} \parallel \hat{c}$ as a function of $\theta$. The colormap represents the normalized oscillator strength, shown on a logarithmic scale. The vertical dashed line defines the angle when the electric field has the direction T$ \rightarrow \Gamma$. The horizontal black line represents the energy of the degenerate exciton $E_{2T}$ which is not affected by the change of direction of the electric field. The dotted crossing curves represent the behaviour of the LE if they were independent, while the magenta and green dotted curves represent the behaviour when they are coupled, according to the two-exciton model in  Appendix \ref{ExcitonsLE}. The exciton energies and oscillator strengths have been obtained by solving BSE for 20 different angles and then interpolated to obtain a smooth function.}
\end{figure}

To perform a more exhaustive analysis of the L-T splitting at $\bf{q}=\Gamma$ of bulk \Bii, Fig.~\hyperref[LTsplitting]{6} shows the energies of the excitons $E_{2-3}$ and $E_{6}$ as a function of the electric field direction. Although the excitons $E_{2-3}$ are degenerate, $E_{2}$ is not influenced by the direction change of the electric field while $E_{3}$ is affected. We have represented the change from in-plane to the out-of-plane direction as a function of
the angle $\theta$, where $\theta=0$ is in-plane and
$\theta=\pi/2$ is out-of-plane direction.
One would expect that the two bright states for the in-plane and out-of-plane case exhibit an independent behaviour, represented by the crossing dashed lines, e.g., the exciton $E_{3}$ would experience the maximum splitting when the electric field direction is in-plane, and its energy would decrease to zero when the electric field direction is out-of-plane, due to the direction-dependent decrease in its oscillator strength. The same trend should be exhibited by the energy shift of $E_{6}$. Instead, the two LE show a coupled behaviour revealed by the avoided-crossing along $\theta$ where one of them has a much lower oscillator strength. This is verified by a two-exciton model, represented in Fig.~\hyperref[LTsplitting]{6} by the magenta and green dashed lines (see Appendix \ref{ExcitonsLE} for a model description of the independent and coupled cases). In essence, the ``transverse'' exciton $E_{6T}$ continuously evolves into $E_{3T}$ without losing its transverse (and optically dark) character, while the ``longitudinal'' split-off exciton $E_{3L}$ becomes the shifted $E_{6L}$ and remains bright for all $\theta$ values. The observation of the L-T splitting as a function of the electric field angle should be possible with electron energy loss spectroscopy (EELS) or inelastic X-ray scattering (IXS).

As a summary, Table~\ref{ExcTransDir} collects the energies of E$_{3}$ and E$_{6}$ for the electric field in-plane and out-of-plane directions. We obtain a value of the L-T splitting of 37 meV for the ${\bf E} \perp \hat{c}$ polarization, excellent agreement with the 38 meV measured experimentally \cite{kaifu1988excitons}. 

\section{Conclusions}\label{discuss}

We have studied the exciton physics of bulk and monolayer \bii with many-body first-principles calculations in the framework of DFT, the GW method and the Bethe-Salpeter equation. We find that both systems exhibit an indirect fundamental gap. The influence of the dimensionality is evident in the electronic and optical properties. The quasiparticle corrections for the monolayer are much larger than for the bulk, which is consistent with the stronger quantum confinement and the reduction of the dielectric screening. The energies of the first bright excitons for ${\bf E} \perp \hat{c}$ and ${\bf E} \parallel \hat{c}$ light polarization in bulk are in good agreement with the experimental values. We demonstrate the large exciton binding energies of bulk and monolayer \Bii, providing a detailed analysis of the excitonic features. The exciton dispersion reaffirms the indirect character of electronic transitions. We find a peculiar direction-dependent hybridization of excitons with different character mediated by the long-range Coulomb interaction embodied in the longitudinal-transverse energy splittings. In addition, our results underscore the importance of a GW+BSE full spinorial description in order to obtain correct quasiparticles and excitons for $\bf{q}=0$ and $\bf{q}\neq0$. Our work provides theoretical support to existing experiment, demonstrate that \bii is an important testbed for the theoretical study of fundamental exciton physics (such as phonon-mediated exciton interaction and localization) and finally confirms that this system is a promising material for the experimental investigation of exciton dynamics (such as tr-ARPES measurements).

\section*{Acknowledgments}

The authors acknowledge the funding of Ministerio de Ciencia e Innovación, which is part of Agencia Estatal de Investigación (AEI), through the project PID2020-112507GB-I00 QUANTA-2DMAT (Novel quantum states in heterostructures of 2D materials) and the Generalitat Valenciana through the Grant PROMETEO/2021/082 (ENIGMA) and the projects SEJIGENT/2021/034 and MFA/2022/009. J. C.-V. acknowledges the Contrato Predoctoral Ref. PRE2021-097581. A. M.-S. acknowledges the Ram\'on y Cajal programme (grant RYC2018-024024-I; MINECO, Spain). This study forms part of the Advanced Materials programme and was supported by MCIN with funding from European Union NextGenerationEU (PRTR-C17.I1) and by Generalitat Valenciana. D.S. and F.P. acknowledge funding from MaX "MAterials design at the eXascale” (Grant Agreement No. 101093374) co-funded by the European High Performance Computing joint Undertaking (JU) and participating countries. D.S. also acknowledges PRIN Grant No. 20173B72NB funded by MIUR (Italy). F.P. acknowledges P. Cudazzo for useful discussions.

\section*{Appendix}

\subsection{Band structure of bulk \Bii}\label{FullBSBiI3}

\begin{figure*}[t!]
\includegraphics[width=0.88\linewidth]{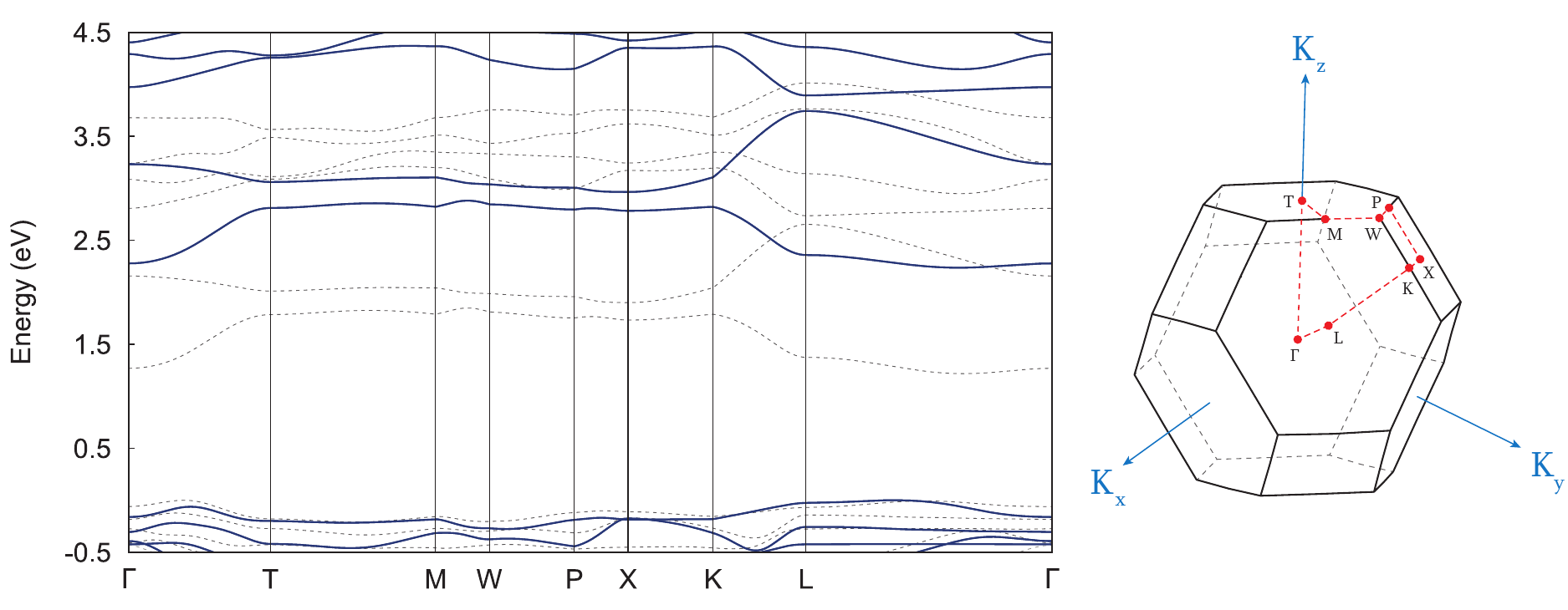}
\caption{\label{BulkFull} DFT and GW band structure along the entire path of the Brilloin zone in bulk \Bii. The solid lines represent the GW bands while the dashed ones represent the DFT ones. The inset shows the high symmetry points in the Brillouin zone.}
\end{figure*}

The Brillouin zone of bulk \bii has multiple high symmetry points to consider, as indicated in Fig. \ref{BulkFull}. However, we have found that not all of these points are relevant in characterising the important features of the band structure.
In order to illustrate the above remark the  band structure of bulk \bii is shown in Fig. \ref{BulkFull} along an extended closed path across the Brillouin zone. We see  that the valence (conduction) band energies 
along the T-M-W-P-X path lie below (above) the corresponding band maximum (minimum), and that their dispersion
is rather flat in that region. This means that at low energies, that portion of the band structure will not significantly contribute to direct and indirect transitions. For this reason and for the sake of clarity, a simplified  path has been used to depict the band structure in  Fig.~\hyperref[Bstructure]{2(a)}.

\begin{figure}[b!]
\includegraphics[width=0.88\linewidth]{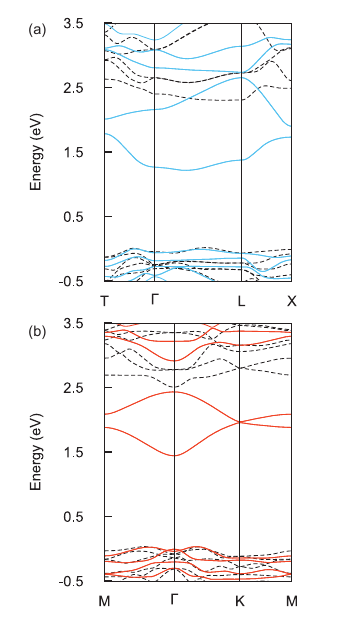}
\caption{\label{BiI3_nosoc} DFT band structure of (a) bulk and (b) monolayer BiI$_{3}$ with (solid lines) and without (dashed lines) spin-orbit coupling.}
\end{figure}

\subsection{Effect of spin-orbit coupling in \Bii}\label{BSnosocBiI3}

The importance of considering the spin-orbit coupling in a material can be mainly predicted depending on the atomic number of the constituent atoms. For large atomic numbers, the effect of the spin-orbit coupling on the material properties can be meaningful. In the case of \Bii, both the bismuth (Z = 83) and iodine (Z = 53) atoms are quite heavy, so that the spin-orbit coupling effect is of vital importance. To illustrate this fact, Fig~\hyperref[BiI3_nosoc]{8(a,b)} shows the DFT band structures for the bulk and monolayer with and without considering the spin-orbit coupling. In both cases the spin-orbit coupling drastically affects the electronic properties by modifying the shape of the bands and shifting down the CBM by 1.0 eV, thus changing the position of the transitions, decreasing the value of the gap and finally modifying the optical properties.

\subsection{Behaviour of L-T splitting as a function of the electric field direction}\label{ExcitonsLE}

The avoided-crossing behaviour manifested by the two LE states in Fig.~\hyperref[LTsplitting]{6} is due to the fact that they are coupled, which can be better understood in terms of a simple two-exciton model. In this model, we reproduce the behaviour of the uncoupled and coupled LE considering explicitly the long-range Coulomb interaction $v^{c}_{G=0}$. We start defining the short-range excitonic Hamiltonian:

\begin{equation}
H^{SR} = H^{0} + V^{SR} - W^{st} 
\end{equation}
where $H^{0}$ is the quasiparticles energy difference (diagonal in the basis of electronic transitions), $V^{SR}$ is the bare electron-hole exchange without including the long-range contribution and $W^{st}$ is the statically screened electron-hole attraction (the term responsible for electron-hole binding), already defined in Eq. \eqref{Eq2}. Then, we define the full excitonic Hamiltonian: 

\begin{equation}
H = H^{SR} + V^{LR}
\label{Eq5}
\end{equation}
with $V^{LR}$ being the long-range exchange contribution (in 3D it originates from the $\mathbf{G}=0$ term of the Coulomb interaction, e.g. $v^{c}_{G=0}$, see discussion in the main text). In the excitonic space ($| \alpha \rangle$, $| \beta \rangle$, being transverse excitonic states which diagonalize $H^{SR}$), for 3D materials and \textbf{q} $\rightarrow$ 0 it reads:

\begin{equation}\label{eq:model}
\langle \alpha | v^{c,LR} | \beta \rangle = C\frac{(\textbf{q} \cdot \textbf{D}^{*}_{\alpha})(\textbf{q} \cdot \textbf{D}_{\beta})}{q^{2}}
\end{equation}
where $C$ is a dimensional constant, \textbf{q} is the momentum vector and $\textbf{D}_{\alpha}$ is the exciton dipole vector, for exciton $\alpha$. We call $E_{b_{\alpha}}$ the eigenvalues of $H^{SR}$ (supposed known from the simulations) and $E_\alpha$ those of $H$, which are to be determined in the model. Since we are dealing with BiI$_3$, we define the bright exciton with dipole along x (in-plane) as E$_{b_{3}}$, and with dipole along z (out-of-plane) as E$_{b_{6}}$. It should be noted that the direction of the excitonic dipoles is also known in advance due to the crystal symmetry.

\vspace{2mm}

\subsection*{Uncoupled bright excitons}

In the case of uncoupled LE the long-range exchange in Eq. \eqref{eq:model} is diagonal and therefore Eq. \eqref{Eq5} gives the following energies:

\begin{equation}
E_{3} = E_{b_{3}} + C|D_{2}|^{2} cos^2\theta
\label{Eq7}
\end{equation}

\begin{equation}
E_{6} = E_{b_{6}} + C|D_{6}|^{2} sin^2\theta
\label{Eq8}
\end{equation}
which is not the coupled behaviour exhibited by the LE in Fig.~\hyperref[LTsplitting]{6} represented by crossed dashed lines.

\vspace{5mm}

\subsection*{Coupled bright excitons}

We now consider that the two excitons $E_{3}$ and $E_{6}$ are coupled. In this situation $H^{SR}_{ij}$ and  $V^{LR}_{ij}$ become $2x2$ matrices and their substitution in Eq. \eqref{Eq5} results in an eigenvalue problem with solutions:

\begin{widetext}
\begin{equation}
E_{\pm} = \frac{E_{b_{3}} + E_{b_{6}} + C|D_{3}|^{2}(a^{2} + b^{2}) \pm \sqrt{ \biggl( C|D_{3}|^{2}(a^{2} + b^{2}) + E_{b_{3}} + E_{b_{6}} \biggr)^{2}  - 4\biggl[ C|D_{3}|^{2}(E_{b_{6}}a^2 + E_{b_{3}}b^2) + E_{b_{3}}E_{b_{6}} \biggr] }}{2}
\label{CoupledExcitons}
\end{equation}
\end{widetext}
where $a = cos\theta$ and $b = \frac{D_{6}}{D_{3}}sin\theta$. In Fig.~\hyperref[LTsplitting]{6} the coupled exciton $E_{+}$ is represented by the magenta curve whereas the other coupled exciton $E_{-}$ is depicted by the green curve. The fact that the solutions of Eq. \eqref{CoupledExcitons} do not perfectly represent the behaviour obtained by BSE is because we are only considering the first bright exciton for the in-plane and out-of-plane directions and therefore we do not consider the effect of other less coupled bright excitons at higher energies. 

%\bibliography{refs}


\end{document}